%% file: main.tex
\documentclass[runningheads]{llncs}
\usepackage{graphicx}

\usepackage{threeparttable}
\usepackage{csquotes}
\usepackage{multirow}
\usepackage{bm}
\usepackage{subcaption}
\usepackage{xspace}
\usepackage{cleveref}
\usepackage{enumitem}
\usepackage{pifont}

\usepackage[font={footnotesize}, skip=2pt]{caption}

\newcommand{\framework}{SENF}
\newcommand{\frameworklong}{Statistical EvaluatioN of Fuzzers}

\begin{document}

\title{My Fuzzer Beats Them All! Developing a Framework for Fair Evaluation and Comparison of Fuzzers}
%
%
\author{David Paaßen\and
Sebastian Surminski\and
Michael Rodler\and
Lucas Davi}
\authorrunning{Paaßen et al.}
\titlerunning{My Fuzzer Beats Them All!}
%
\institute{University of Duisburg-Essen, Germany
\email{\{david.paassen,sebastian.surminski,michael.rodler,lucas.davi@uni-due.de\}}}
\maketitle              

\begin{abstract}
\input{00-abstract}
\end{abstract}

\input{01-introduction}
\input{02-background}
\input{03-statistical_evaluation}
\input{04-problem_description}
\input{05-methodology}
\input{06-experiments_extended}
\input{07-discussion_extended}
\input{08-conclusion}

\input{09-acknowledgements}
%
%
%
\bibliographystyle{splncs04}
\bibliography{references}

\input{10-appendix}
\end{document}

%% file: 00-abstract.tex
Fuzzing has become one of the most popular techniques to identify bugs in software. To improve the fuzzing process, a plethora of techniques have recently appeared in academic literature. However, evaluating and comparing these techniques is challenging as fuzzers depend on randomness when generating test inputs. Commonly, existing evaluations only partially follow best practices for fuzzing evaluations. We argue that the reason for this are twofold. First, it is unclear if the proposed guidelines are necessary due to the lack of comprehensive empirical data in the case of fuzz testing. Second, there does not yet exist a framework that integrates statistical evaluation techniques to enable fair comparison of fuzzers.

To address these limitations, we introduce a novel fuzzing evaluation framework called \framework~(\frameworklong). We demonstrate the practical applicability of our framework by utilizing the most wide-spread fuzzer AFL as our baseline fuzzer and exploring the impact of different evaluation parameters (e.g., the number of repetitions or run-time), compilers, seeds, and fuzzing strategies. Using our evaluation framework, we show that supposedly small changes of the parameters can have a major influence on the measured performance of a fuzzer.

%% file: 01-introduction.tex
\section{Introduction}
\label{sec:introduction}
Fuzzing approaches aim at automatically generating program input to assess the robustness of a program to arbitrary input. The goal of a fuzzer is to trigger some form of unwanted behavior, e.g., a crash or exception. Once a program fault occurs during the fuzzing process, a developer or analyst investigates the fault to identify its root cause.
Subsequently, this allows the software vendor to improve the quality and security of the software.
One of the most prominent fuzzers, called American Fuzzy Lop (AFL)~\cite{ZAL2019}, has discovered hundreds of security-critical bugs in a wide variety of libraries and programs. 

Following the success of AFL, various other fuzzers have been proposed which aim to outperform AFL by implementing new and improved fuzzing techniques (e.g.,~\cite{LEM2018,BOE2016,LYU2019,GAN2018}). However, it remains largely unclear whether the claim of improving the overall fuzzing performance is indeed true. This is because accurately evaluating a fuzzer is challenging as the fuzzing process itself is non-deterministic. Hence, comparing single runs or multiple runs using simple statistical measurements such as average values can lead to false conclusions about the performance of the evaluated fuzzer. Similarly, deriving the number of potentially discovered bugs based solely on coverage measurements and the number of program crashes does not necessarily map to the effectiveness of a fuzzer. For instance, Inozemtseva et al.~\cite{INO2014} show that there is no strong correlation between the coverage of a test suite and its ability to detect bugs. Additionally, there are fuzzing approaches that prioritize certain program paths instead of maximizing the overall coverage~\cite{BOE2017,HON2018,WAN2020}. Such approaches cannot be evaluated using overall code coverage as a measurement.

A study by Klees et al.~\cite{KLE2018} shows that existing evaluation strategies do not consider state-of-the-art best practices for testing randomized algorithms such as significance tests or standardized effect sizes. They also provide a list of recommendations. However, these recommendations are mainly derived from known best practices from the field of software testing or from a small set of experiments on a small test set. Nevertheless, as we will show in Section~\ref{sec:problem_description}, recent fuzzing proposals still do not consistently follow recommendations regarding the employed statistical methods and evaluation parameters (e.g., run-time or number of trials). Since the goal of the recommendations is to ensure that the reported findings are not the results of randomness, it remains unclear whether we can trust existing fuzzing experiments and conclusions drawn from those experiments. 

Another important aspect of any fuzzer evaluation concerns the employed test set. Several research works introduced test sets such as LAVA-M~\cite{DOL2016}, Magma~\cite{HAZ2020}, or the Google Fuzzer Suite~\cite{GOO2016}. Ideally, a test set should contain a wide variety of different programs as well as a set of known bugs covering various bug types including a proof-of-vulnerability (PoV). This is crucial to enable accurate assessment on the effectiveness and efficiency of a fuzzer as a missing ground truth may lead to overestimating or underestimating the real performance of a fuzzer. We analyze these test sets in detail in Section~\ref{subsec:methodology:testsets} as the test set selection is crucial for evaluating and comparing fuzzers.

Lastly, existing evaluation strategies lack uniformity for evaluation parameters such as the number of trials, run-time, and size of the employed test set and the included bugs. As it is still unknown how these parameters affect the fuzzer evaluation in practice, fuzzing experiments are commonly executed using a wide variety of different parameters and evaluation methods. This may not only affect the soundness (e.g., due to biases caused by the choice of parameter) of the results but also makes it even harder to compare results across multiple studies.

\smallskip
\noindent\textbf{Our Contributions.}
In this study, we address the existing shortcomings of fuzzing evaluations. To do so, we review current fuzzing evaluation strategies and introduce the design and implementation of a novel fuzzing evaluation framework, called \framework~(\frameworklong), which unifies state-of-the-art statistical methods and combines them to calculate a ranking to compare an arbitrary number of fuzzers on a large test set. The goal of our framework is twofold. First, we aim to provide a platform that allows us to easily compare a large number of fuzzers (and configurations) on a test set utilizing statistical significance tests and standardized effect sizes. Contrary to existing frameworks, such as UNIFUZZ~\cite{LI2021}, \framework\ provides a single ranking which allows for an easy comparison of the overall performance of fuzzers. Second, due to the lack of comprehensive empirical data we test if following the recommended best practices is necessary or if we can loosen the strict guidelines to reduce the computational effort needed to compare different fuzzing algorithms without impairing the quality of the evaluation which was not possible with the data provided by Klees et al.~\cite{KLE2018}.

To show the applicability of \framework, we build our evaluation based on the most prominent fuzzer, namely AFL~\cite{ZAL2019} and its optimizations, as well as the popular forks AFLFast~\cite{BOE2016}, Fairfuzz~\cite{LEM2018}, and AFL++~\cite{FIO2020}. This allows us to argue about the usefulness and impact of the proposed methods and techniques as AFL is commonly used as the baseline fuzzer in existing fuzzing evaluations. We ensure that all tested fuzzers share the same code base which allows us to precisely attribute performance differences to the changes made by the respective fuzzer or optimization technique.

We provide an extensive empirical evaluation of the impact of fuzzing parameters. In total, we ran over 600 experiments which took over 280k CPU hours to complete. To the best of our knowledge, this is currently the largest study of fuzzers published in academic research.

In summary, we provide the following contributions:
\begin{itemize}
	\item We implement a fuzzing evaluation framework, called \framework, which utilizes state-of-the-art statistical evaluation methods including p-values and standardized effect sizes to compare fuzzers on large test sets.
	\item We conduct a large-scale fuzzer evaluation based on a test set of 42 different targets with bugs from various bug classes and a known ground truth to quantify the influence of various evaluation parameters to further improve future fuzzer evaluations.
	\item We open-source \framework~\cite{PAA2021}, containing the statistical evaluation scripts, the result data of our experiments, and seed files to aid researchers to conduct fuzzer evaluations and allowing reproducibility of our study.
\end{itemize}

%% file: 02-background.tex
\section{Background}
\label{sec:background}
In this section, we provide background information on the most relevant fuzzing concepts and discuss how these are implemented in case of the popular fuzzer AFL~\cite{ZAL2019}.

Fuzzers are programs that need to decide on a strategy to generate inputs for test programs. The inputs should be chosen in such a way that they achieve as much coverage of the program's state space as possible to be able to detect abnormal behavior that indicates an error. Fuzzers are commonly categorized into black-box, white-box, and grey-box fuzzers. 
Where black-bock fuzzers (e.g., zzuf~\cite{HOV2006}) try to maximizes the number of executions while white-box fuzzers (e.g., KLEE~\cite{CAD2008}) make heavy use of instrumentation and code analysis to generate more significant input. Grey-box fuzzers (e.g., AFL~\cite{ZAL2019}) try to find a balance between the executions per second and time spend on analysis.

One of the most well-known fuzzers is called American fuzzy lop (AFL) and is a mutation-based coverage-guided grey-box fuzzer. It retrieves coverage feedback about the executed program path for a corresponding test input. Since its creation, AFL discovered bugs in more than 100 different programs and libraries~\cite{ZAL2019} confirming its high practical relevance to improve software quality and security.

Given the influence of AFL in the fuzzing area, we take a closer look at its architecture.
AFL includes a set of tools that act as drop-in replacements for known compilers, e.g., as a replacement for \texttt{gcc} AFL features \texttt{afl-gcc} which is used to add code instrumentation. The instrumentation provides crucial information such as the branch coverage and coarse-grained information about how often a specific branch has been taken.

\begin{figure}[htbp]
\center
\includegraphics[width=0.6\columnwidth]{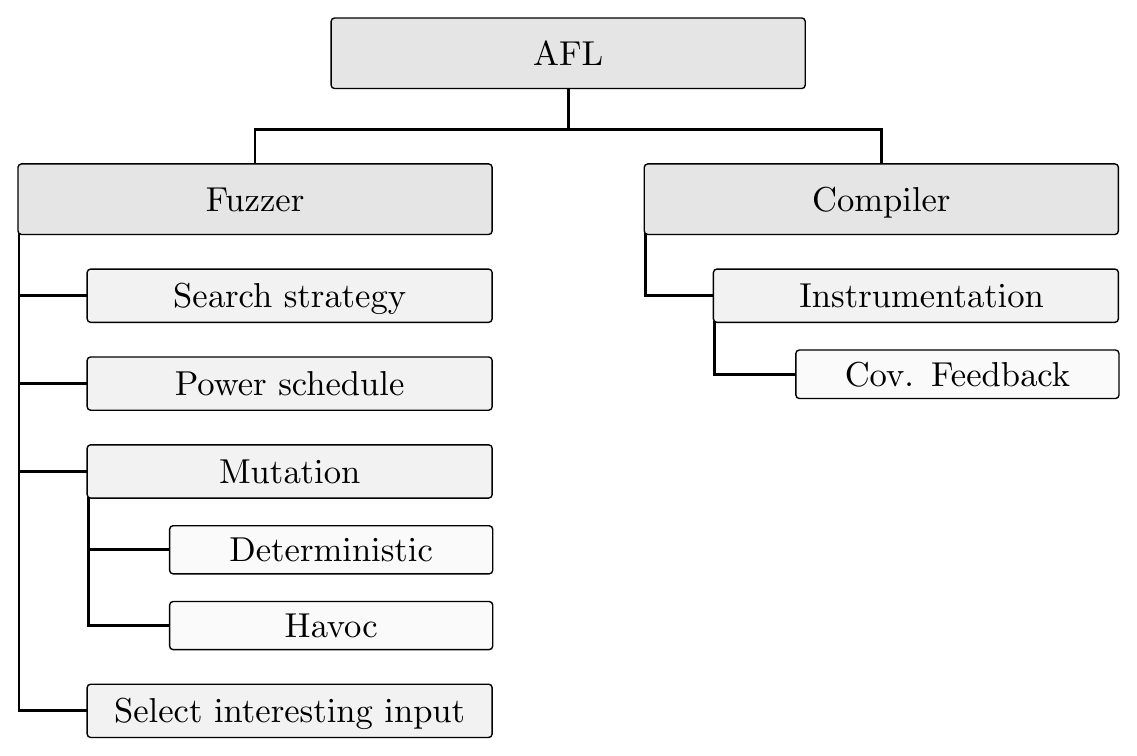}
\caption{Overview of the different components of AFL.}
\label{fig:afloverview}
\end{figure}

The fuzzing process can be divided into four different core components (see Figure~\ref{fig:afloverview}) which can also be found in many existing mutations-based grey-box fuzzers, including forks of AFL:
\ding{172}~\emph{Search strategy}: The search strategy selects an input (e.g., one of the initial seeds) that is used in the mutation stage to generate more test inputs. \ding{173}~\emph{Power schedule}: The power schedule assigns an energy value which limits the number of new inputs generated in the mutation stage. The idea is to spend more time mutating input that is more likely to increase the code coverage.  \ding{174}~\emph{Mutations}: The mutation stage changes (part of) the selected input to produce new inputs which are then executed by the program under test. AFL has two different mutation stages. The \emph{deterministic stage} does simple bit flips or inserts specific values such as \texttt{INT\_MAX}. In the \emph{havoc stage}, AFL executes a loop that applies different mutations on the selected input, e.g., inserting random data or trimming the input.   \ding{175}~\emph{Select interesting input}: After executing a new input, the fuzzer collects the feedback data and decides if the newly generated input is interesting, i.e., whether or not the input should be mutated to generate new inputs.

Successors of AFL commonly implement their improvements as part of one or more of the discussed core components. In Section~\ref{subsec:experiments:fuzzers} we describe the changes implemented by the different fuzzers we test in our evaluation in more detail.

To address the problem of inputs being rejected due to rigorous input checks, fuzzer leverage seed files which provide initial coverage and useful inputs for the mutation stage. Thus, a fuzzer does not need to learn the input format from scratch. To generate a set of seed files, one can either collect sample files from the public sources or manually construct them. AFL prefers seeds with high code coverage, a small file size, and low execution time. To minimize the seed set, AFL provides a tool called \texttt{afl-cmin} which one can use to remove useless seed files. However, if it is not possible to collect a sophisticated seed set one can always employ an empty file as the only initial seed file.

%% file: 03-statistical_evaluation.tex
\section{Statistical Evaluations}
\label{subsec:backgroud:statisticalevaluation}

As the main purpose of fuzzers is to find bugs, the naive approach to compare two or more fuzzers, is to fuzz a target program for a fixed amount of time and then either compare the time it took to find bugs or compare the number of bugs a fuzzer discovered. However, the fuzzing process itself is non-deterministic. For instance, in AFL, the non-deterministic component is the havoc stage which is part of the mutation module. Thus, executing multiple trials with the same fuzzer may yield different results. As a consequence, using only a single execution might lead to a false conclusion. Other utilized evaluation metrics such as the average and median can be affected by similar issues. The common problem of these simple techniques is that they ignore randomness, i.e., they do not consider the non-deterministic nature of fuzzing. The most common method to address this problem is to calculate the statistical significance, i.e., the p-value which was popularized by Fisher~\cite{FIS1925}. If the p-value is below a predefined threshold we assume that the observed difference between to fuzzers is genuine and consider the results statistically significant.

When comparing two fuzzers, it is not only relevant to know whether the performance differences are statistically significant but also to properly quantify the difference, namely, we have to calculate the effect size. However, when comparing fuzzers on multiple targets non-standardized effect sizes are affected by the unit of measurement which may result in unwanted biases. To address this issue a standardized effect size should be used~\cite{ARC2014}. 

In general, we can differentiate between statistical tests for dichotomous and interval-scale results which require a different set of statistical evaluation methods. In the following, we describe both result types and the recommended approach to calculate statistical significance and the corresponding effect size as discussed by Arcuri et al.~\cite{ARC2014}. For more details about the employed statistical methods, we refer the interested reader to the relevant literature~\cite{MAN47,FIS22,VAR00}.

An interval-scale result in the context of fuzzing is the time a fuzzer needs to detect a specific bug. In such a case it is recommended to use the Mann-Whitney-U test to calculate the p-value to test for statistical significance. Contrary to the popular \emph{t-test} the Mann-Whitney-U test does not assume that the underlying data follows the normal distribution. To quantify the effect size for interval-scale results, one can utilize the Vargha and Delaneys $\hat{A}_{12}$ statistic.

A dichotomous result can only have two outcomes, usually \emph{success} or \emph{failure}. In the context of a fuzzer evaluation, a dichotomous result simply states whether a specific bug has been discovered in the given time limit. To calculate the statistical significance, Arcuri et al.~\cite{ARC2014} recommend using the Fisher exact test. As the name suggests, this statistical test is exact which means that it is precise and not just an estimation for the actual p-value. To calculate the effect size for dichotomous results, it is recommended to calculate the odds ratio.

%% file: 04-problem_description.tex
\section{Problem Description and Related Work}
\label{sec:problem_description}

The evaluation of fuzzers was first analyzed by Klees et al.~\cite{KLE2018} who demonstrate that simply comparing the number of crashes found using a single trial on a small set of targets is misleading as it gives no insight into whether the fuzzer finding more crashes discovers more bugs in practice. Thus, it is preferred to use a test set with a ground truth, i.e., a set of inputs that trigger a known bug or vulnerability inside the test program. To improve fuzzer evaluations, Klees et al.~\cite{KLE2018} provided a set of recommendations for evaluating fuzzers based on best practices from the field of software engineering. They recommend 30~trials, a run-time of 24h and use of the Mann-Whitney-U test for statistical significance, and the $\hat{A}_{12}$ statistic as an effect size. However, as we show in Table~\ref{tab:evalsummery}, these recommendations are only partially followed by current fuzzing evaluations. As it is unknown how much influence each evaluation parameter has on the results, it is unclear whether or not these results are reproducible in practice. Contrary to Klees et al~\cite{KLE2018}, we conduct comprehensive experiments to be able to argue about the influence of different evaluation parameters based on empirical data. 

To discuss the state of current fuzzer evaluations we analyze the evaluations from previous work published in reputable security conferences. The experiments gathered from the evaluation sections of different studies based on the following criteria: \ding{172}~the experiment is used to compare the overall performance of the respective approach to at least one different fuzzers \ding{173}~we exclude experiments that are used to either motivate the work or certain design choices as well as case studies. The results are summarized in Table~\ref{tab:evalsummery}. Note that we use the term \emph{Crashes} as an evaluation metric for all evaluations that do not utilize a ground truth and rely on a de-duplication method which tries to correlate crashes to a root cause. However, de-duplication methods are prone to errors and cannot sufficiently estimate the correct number of bugs~\cite{KLE2018}. We use the term \emph{Bugs} when the authors evaluate fuzzers with a set of targets that contain known vulnerabilities, i.e., it is possible to determine which inputs trigger which bug without utilizing a de-duplication technique.

We observe that none of the fuzzing proposals strictly follows all best practices in their evaluations. For instance, none of the listed studies uses 30 trials per experiment and only a single study employs a standardized effect size. Another problem is the lack of uniformity. This is especially prevalent when real-world programs are used to evaluate fuzzers which regularly use different sets of programs or program versions which may introduce unwanted biases and also makes it hard to compare these results. Furthermore, most studies either do not provide any statistical significance results or only for some of the conducted experiments.

\begin{table*}[htbp]
\let\TPToverlap=\TPTrlap
\centering
\caption{\footnotesize Analysis of current fuzzer evaluations. Entries with a question mark mean that we were unable to find the respective information in the related study. Test set: \emph{RW} = real-world programs, \emph{Google} = Google fuzzing suite. The number following the test sets corresponds to the number of targets used. Effect Size: \emph{Avg.} = average, \emph{Max.} = maximum value of all trials. Statistical significance: \emph{CI} = confidence intervals, \emph{MWU} = Mann-Whitney-U test.\\}
\begin{threeparttable}
\resizebox{1.0\textwidth}{!}{
\begin{tabular}{lcllllll}
\hline
Fuzzer & No. & Test set & Run-time & Trials & Eval. metric & Effect size & Stat. significance \\ 
\hline 
\multirow{4}{*}{Hawkeye~\cite{HON2018}} & 1 & RW (19) & 8h & 20 & Bugs & Average & - \\ 
 & 2 & RW (1) & 4h & 8 & Bugs & Average, $\hat{A}_{12}$ & - \\ 
& 3 & RW (1) & 4h & 8 & Bugs & Average, $\hat{A}_{12}$ & - \\ 
& 4 & Google (3) & 4h & 8 & Coverage & Average, $\hat{A}_{12}$ & - \\
\hline
\multirow{3}{*}{Intriguer~\cite{CHO2019}} & 1 & LAVA-M (3) & 5h & 20 & Bugs & Median, Max. & \\
 & 2 & LAVA-M (1) & 24h & 20 & Bugs & Median & CI, MWU\\ 
 & 3 & RW (7) & 24h & 20 & Coverage & Median &CI\tnote{2}, MWU\\
 \hline
\multirow{4}{*}{DigFuzz~\cite{ZHA2019}} & 1 &CGC (126) & 12h & 3 & Coverage & Norm. Bitmap\tnote{1} & - \\
 & 2 & CGC (126) & 12h & 3 & Bugs & - & - \\ 
 & 3 & LAVA-M (4) & 5h & 3 & Bugs & ? & - \\ 
& 4 & LAVA-M (4) & 5h & 3 & Coverage & Norm. Bitmap\tnote{1} & - \\
 \hline
\multirow{4}{*}{REDQUEEN~\cite{ASH2019}} & 1 & LAVA-M (4) & 5h & 5 & Bugs & Median & CI\tnote{3}\\ 
 & 2 & CGC (54) & 6h & ? & Bugs & - & - \\ 
 & 3 & RW (8) & 10h & 5 & Coverage & Median & CI, MWU\\
  & 4 & RW (8) & 10h & 5 & Bugs & - & -\\
 \hline
 \multirow{4}{*}{GRIMOIRE~\cite{BLA2019}} & 1 & RW (8) & 48h & 12 & Coverage & Median & CI, MWU\tnote{4}\\ 
 & 2 & RW (4) & 48h & 12 & Coverage & Median & CI, MWU\\ 
 & 3 & RW (3) & 48h & 12 & Coverage & Median & CI, MWU\\ 
 & 4 & RW (5) & ? & ? & Bugs & - & - \\ 
 \hline
 \multirow{4}{*}{EcoFuzz~\cite{YUE2020}} & 1 & RW (14) & 24h & 5 & Coverage & Average & p-value\tnote{5} \\
 & 2 & RW (2) & 24h & 5 & Coverage & Average & - \\
 & 3 & RW (2) & 24h & ? & Crashes & - & - \\
  & 4 &  LAVA-M (4) & 5h & 5 & Bugs & - & - \\
 \hline
\multirow{5}{*}{GREYONE~\cite{GAN2020}} & 1 & RW (19) & 60h & 5 & Crashes\tnote{6} & - & - \\
& 2 & RW (19) & 60h & 5 & Coverage & - & - \\
& 3 & LAVA-M (4) & 24h & 5 & Bugs & Average & - \\
& 4 & LAVA-M (4) & 24h & 5 & Crashes & Average & - \\
& 5 & RW (10) & 60h & 5 & Coverage & Average& - \\
\hline
\multirow{3}{*}{Pangolin~\cite{HUA2020}} & 1 & LAVA-M (4) & 24h & 10 & Bug & Average & MWU \\
& 2 & RW (9) & 24h & 10 & Crashes & - & - \\
& 2 & RW (9) & 24h & 10 & Coverage & Average & MWU \\
\hline
\end{tabular}}
\begin{tablenotes}\scriptsize
\item [1] Normalized Bitmap size describes the relative size of the bitmap compared to the bitmap found\\ by all tested fuzzers.
\item [2] Confidence intervals only given for five of the seven targets.
\item [3] Confidence intervals are only provided for Redqueen.
\item [4] The Appendix further provides: mean, standard deviation, skeweness, and kurtosis.
\item [5] We were unable to determine the exact statistical test which has been used to obtain the p-value.
\item [6] The evaluation compares de-duplicated crashes as well as unique crashes as reported by\\ AFL-style fuzzers.
\end{tablenotes}
\end{threeparttable}
\label{tab:evalsummery}
\end{table*}

A work that partially addresses similar issues has been introduced by~Metzman et al.~\cite{MET2020} from Google who published FuzzBench, an evaluation framework for fuzzers. FuzzBench generates a report based on coverage as an evaluation metric including a statistical evaluation. However, as the main purpose of a fuzzer is to find bugs, the coverage is only a proxy metric and therefore less meaningful than comparing the number of bugs found on a ground truth test set and thus not recommended~\cite{PHA2017,VTO2018,KLE2018}.

UNIFUZZ is a platform to compare different fuzzers based on 20 real-world programs~\cite{LI2021}. The evaluation metrics are based on crashes which are de-duplicated using the last three stack frames which is known to be unreliable because stack frames might be identical even though they trigger different bugs or stack frame may be different while triggering the same bug~\cite{KLE2018}. UNIFUZZ provides an overview of the fuzzer performance on each test program which makes it hard to assess the overall performance. \framework\ goes one step further and summarizes the results in a single ranking which allows us to easily compare all tested fuzzers. Therefore, it is not required to manually analyze the results on each target separately. However, if needed one can still get the target specific data from the results database of \framework.

%% file: 05-methodology.tex
\section{Our Methodology}
\label{sec:methodology}
In this section, we provide an overview of the most important aspects of a fuzzer evaluation. We describe our choice of fuzzers, seeds, test set, and test machine setup which we use to test our framework to quantify the influence of various evaluation parameters.

\subsection{Comparing Fuzzers}
\label{subsec:methodology:fuzzers}

Comparing fuzzers with each other is not straightforward due to the various fuzzer designs and the wide variety of available testing methods. A fuzzer design is usually highly complex and given that a fuzzer executes millions of test runs, even small differences can have a huge impact on the performance. Some fuzzers are based on completely novel designs which makes it hard to attribute performance improvements to a specific change. For example, Chen and Chen proposed Angora~\cite{CHE2018} a mutation-based fuzzer that is written in Rust instead of C/C++ like AFL. Angora implements various methods to improve the fuzzing process: byte-level taint tracking, a numeric approximation based gradient descent, input length exploration, and integration of call stacks to improve coverage mapping. Due to the considerable differences to other fuzzers, it is impossible to accurately quantify the respective contribution of each technique when comparing it with AFL or other fuzzer which do not share the same code base. As a consequence, it is important to evaluate fuzzers based on common ground. Given the high popularity of AFL, we opted to focus on fuzzers that are based on the AFL code base. Note however that our evaluation framework is not specifically tailored to AFL in any way. Thus, it can be used to evaluate an arbitrary selection of fuzzers.

\subsection{Test Set Selection}
\label{subsec:methodology:testsets}
A crucial aspect of any fuzzer evaluation is the underlying test set, i.e., the target programs for which the fuzzer aims to discover bugs. In what follows, we study four different test sets available at the time of testing and argue why we decided to use the CGC test set for our evaluation. Note that we focus on test sets that provide a form of ground truth as there is currently no way to reliably match crashes to the same root cause as proper crash de-duplication is still an open problem (see Section~\ref{sec:problem_description}).

\smallskip
\noindent\textbf{LAVA-M.}
In 2016, Brendan et al. presented LAVA~\cite{DOL2016}, a method to inject artificial bugs into arbitrary programs. The corresponding \emph{LAVA-M} test set was the first ground truth test set to evaluate fuzzers that has been published in academia. It consists of four different programs with hundreds of injected bugs. Each bug has a specific bug-id that is printed before deliberately crashing the program. Due to its rather small size, the \emph{LAVA-M} test set lacks the diversity found in real-world programs. Further, recent fuzzers such as Redqueen~\cite{ASH2019} and Angora~\cite{CHE2018} solve the test set by finding all injected bugs. This is possible because LAVA-M only features a single bug type which requires that the fuzzer solves magic byte comparisons, missing the bug diversity found in real-world software.

\noindent\textbf{Google Fuzzing Suite.}
The Google Fuzzer Suite~\cite{GOO2016} consists of 22~different real-world programs with 25 different challenges that fuzzers are expected to solve. All challenges are documented and may include seed files. However, the test~suite is not suitable for our use case as the majority of the bugs can be discovered by existing fuzzers in a very short time span (seconds or minutes). Furthermore, some of the challenges do not contain any bugs. Instead, the goal of these challenges is for the fuzzer to reach a certain path or line of code (i.e., a coverage benchmark) which is not compatible with our evaluation metric as we are interested in the number of bugs found. Additionally, the included bugs are partially collected from other fuzzing campaigns which might introduce biases.

\noindent\textbf{Magma.}
The Magma fuzzing benchmark is a ground truth test set~\cite{HAZ2020} that is based on a set of real-world programs. At the time of testing, the test set contains six different targets each containing a set of known bugs. Similar to LAVA-M, Magma uses additional instrumentation in the form of bug oracles to signal whether a bug condition has been triggered by a specific input.

However, we do not use the Magma test set because at the time of testing it did not provide a sufficiently large test set. Further, not all bugs include a proof-of-vulnerability (PoV) which makes it impossible to know if a fault can be triggered by any means.

\noindent\textbf{CGC.}
The DARPA Cyber Grand Challenge\footnote{\url{https://github.com/CyberGrandChallenge/}} (CGC) was a capture-the-flag style event where different teams competed by writing tools that are able to detect and subsequently fix bugs in a test corpus of close to 300 different programs with a prize pool of nearly 4 million USD. Each challenge has been designed carefully and consists of one or more binary which mirror functionality known from real-world software. CGC challenges contain at least one bug of one of two types: Type 1 bugs allow an attacker to control the instruction pointer and at least one register. Type 2 bugs allow reading sensitive data such as passwords. The challenges are written by different teams of programmers and do not rely on automatically injected bugs. As a result, the CGC test set offers great bug diversity which are similar to bugs found in real-world software and is therefore not susceptible to the same limitations as the LAVA-M test set.

Instead of the original binaries which were written for \emph{DECREE}, we use the multi OS variant published by Trail of Bits~\cite{GUI2016} which allows us to execute the challenge binaries on Linux. All challenges and bugs are very-well documented and contain a PoV and a patched version of the respective challenge program(s). Each bug is categorized into their respective CWE classes\footnote{Common Weakness Enumeration (CWE) is a list of software and hardware problem types (\url{https://cwe.mitre.org/})}. Further, challenges include test suits that we can use to ensure that the compiled program works as intended which can be especially helpful when using code instrumentation. 

Given the greater bug and program diversity of CGC in combination with its great documentation and comprehensive test~suites, we select a subset of the ported version of the CGC test set based on the following criteria: \ding{172}~All tests (including the PoV) are successfully executed on our test servers. \ding{173}~The target only contains one vulnerability. \ding{174}~The vulnerability is of type 1 as type 2 bugs do not lead to a crash. \ding{175}~The challenge consists of only one binary as fuzzers usually do not support to fuzz multiple binaries.

We are using targets with only one vulnerability as this allows us to verify the discovered crashing inputs using differential testing (see Section~\ref{subsec:methodology:testruns}). This process does not require any additional instrumentation (e.g., bug oracles) which may significantly change the program behavior and lead to non-reproducible bugs~\cite{LI2021}. Furthermore, we do not need to correlate crashes to their root cause using de-duplication methods. Our complete test set is composed of 42 targets including bugs of 21 different CWE types. We provide a complete list of all targets including their bug types in Appendix~\ref{apx:subsec:testset}.
Note that it is not required to use CGC to be able to use \framework\ because the framework is not specifically tailored to the test set but can used with any test set.

\subsection{Seed sets}
\label{subsec:methodology:seeds}

To evaluate fuzzers, we opted to use two sets of seed files. The first set of seed files contains sample input which we extract from the test inputs that are shipped with each CGC challenge. We minimize each seed set using \texttt{afl-cmin}. As it might not always be possible for users to create a comprehensive seed set for their target, we use an empty file as a second seed set.

\subsection{Statistical Evaluation}
\label{subsec:methodology:statisticalevaluation}

To evaluate the results of our experiments, we employ the statistical methods described in Section~\ref{subsec:backgroud:statisticalevaluation}. \framework\ supports both, dichotomous and interval-scale statistics as their usage depends on the use case. Dichotomous results provide an answer to the question which fuzzer finds the most bugs in a certain time frame, but ignores the time it took to find a bug. These types of evaluations are relevant for use cases such as OSS-Fuzz~\cite{AIZ2016} where fuzzing campaigns are continuously run for months without a fixed time frame. Statistical tests on interval-scale results are useful in practical deployments when the amount of time to fuzz a target is limited, e.g., when running tests before releasing a new software version. We use R~\cite{RCT2019} to calculate statistical significance tests as well as effect sizes.

When comparing multiple fuzzers or fuzzer configurations on a large set of targets, we encounter two problems. First, due to the large number of comparisons, it is not practical to publish all p-values and effect sizes as part of a study. Secondly, even if one publishes all values, it is not trivial to assess if a fuzzer actually outperforms another. Therefore, we implement a scoring system, which is inspired by Arcuri et al.~\cite{ARC2014}, to summarize the results in a single score. The scoring system follows the intuition that the fuzzer which finds the most bugs the fastest, on the most of the targets is overall the best fuzzer. To determine the best fuzzer, the algorithm compares all fuzzers using the statistical significance tests and standardized effect sizes. For each target, it generates a ranking based on the time it took each fuzzer to find a specific bug. The final score is the average ranking of each fuzzer over the whole test set. For a more detailed description of the scoring algorithm we refer the interested reader to the respective publication~\cite{ARC2014}.

\subsection{Fuzzing Evaluation Setup}
\label{subsec:methodology:framework}

\begin{figure}[htb]
\center
\includegraphics[width=0.95\textwidth]{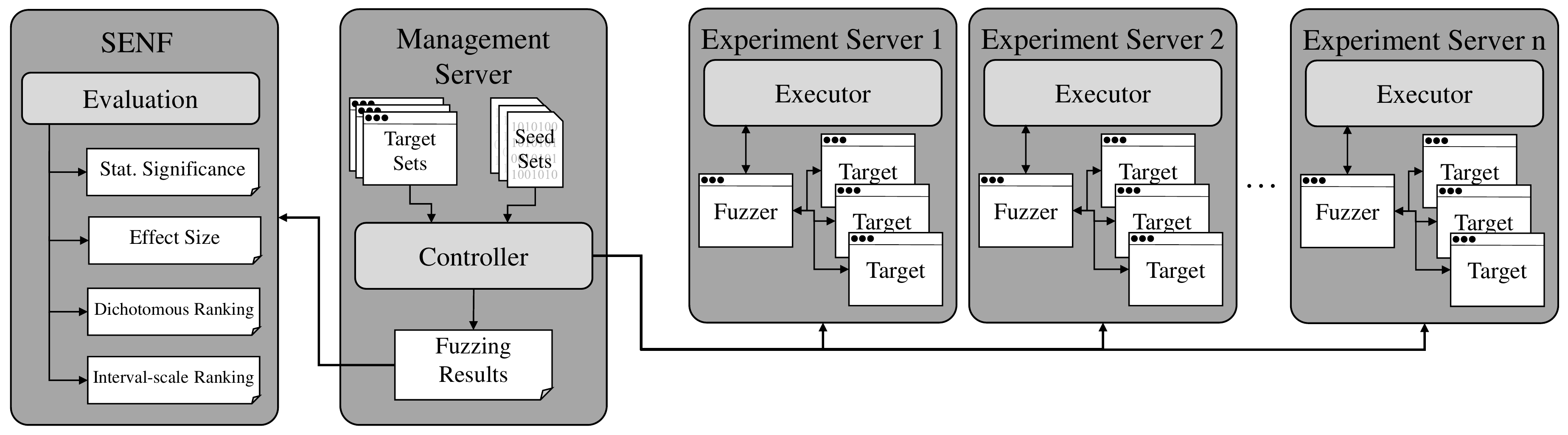}
\caption{Conceptual overview of the fuzzing evaluation setup.}
\label{fig:approachoverview}
\end{figure}

In Figure~\ref{fig:approachoverview} we provide an overview of our fuzzing evaluation setup. At its core, our design features a management server that runs a controller which provides the target program and seed set to one of the available experiment servers. On each experiment server, a dedicated executor starts the fuzzer and monitors the fuzzing process. The monitoring includes logging the CPU utilization and number of executions of the fuzzer. Thus, we can detect hangs and crashes of the fuzzer itself and restart a run if necessary.
After the pre-defined run-time, the executor stops the fuzzer and sends a success message to the controller program. Using the data from all successful fuzzing runs, \framework\ evaluates the reported results using evaluation methods which compare all executed runs of an arbitrary number of fuzzers and calculates statistical significance, effect size, the ranking based on dichotomous and interval-scale statistical tests.

\subsection{Test runs}
\label{subsec:methodology:testruns}

Note that we conduct each experiment (i.e., a combination of fuzzers, targets, and seeds) for a maximum of 24h. As each target only contains a single bug, we stop fuzzing when an input has been found by a fuzzer that triggers the bug of the respective target. To avoid false positives~\cite{ASH2019}, we verify each crash using differential analysis, i.e., we execute a potential crashing input on the vulnerable and patched version of the respective binary and check whether the input crashes the binary.

\framework\ itself only requires a database which contains the result data, i.e., the time it took until a specific bus has been triggered, and a list of targets/seeds that should be used in the evaluation. Therefore, our framework can be used to evaluate other fuzzers or test sets. With minimal modifications one can also use other evaluation metrics (e.g., block coverage) to compare fuzzers with \framework.

%% file: 06-experiments_extended.tex
\section{Experiments}
\label{sec:experiments}

We run extensive fuzzing campaigns to systematically quantify the influence of various parameters used in fuzzing evaluations while following state-of-the-art statistical evaluation methodology. We test the influence of the following parameters: the seed set, number of trials, run-time, and number of targets as well as bugs. In total we run 616 fuzzing experiments with an accumulated run-time of over 284k CPU hours.

If not stated otherwise we use the following parameters as a default configuration for the statistical evaluation: a p~threshold of 0.05, a non-empty seed set, interval-scale statistical tests, with 30 trials per experiment and a run-time of 24h.
Further, experiments for a specific target are always run on the same hardware configuration to ensure uniform test conditions. Note that when testing with an empty seed we have to exclude seven targets of our test set of 42 programs as these targets do not properly process an empty file thus fail initial tests done in AFLs initialization routine.

We execute all experiments on a cluster of 13 servers. To ensure equal conditions for all fuzzers, we use Docker containers and assign them one CPU core each and a ramdisk to minimize the overhead caused by I/O operations. We utilize Ubuntu 18.04 LTS as an operating system. If not stated otherwise we use fuzzers AFL/Fairfuzz in version 2.52b and AFLFast in version 2.51b as well as AFL++ in version 2.65c. The CGC test set was built using the code from commit \texttt{e50a030} from the respective repository from Trail of Bits.

\subsection{Fuzzers}
\label{subsec:experiments:fuzzers}

We test a total of four fuzzers (\ding{182}~AFL~\cite{ZAL2019}, \ding{183}~AFLFast~\cite{BOE2016}, \ding{184}~Fairfuzz~\cite{LEM2018}, \ding{185}~AFL++~\cite{FIO2020}), two AFL-based compiler optimizations (\ding{186}~AFL-CF, \ding{187}~AFL-LAF), and two modes of AFL  (\ding{188}~\texttt{-d} and \ding{189}~\texttt{-q}) which provide a wide range of different performances. In the following, we explain the different fuzzers and modes of AFL we tested including the different compiler optimizations. 

\noindent
\ding{182}~\textbf{AFL.} The general purpose fuzzer AFL supports various different optimizations and parameters which change one or more its core components:
\ding{188}~\textbf{AFL (-d).} If the \texttt{-d} flag is enabled, AFL skips the deterministic part of the mutation stage and directly proceeds with the havoc stage. \ding{189}~\textbf{AFL (-q)} The \texttt{-q} flag enables the \emph{qemu mode}. Using this mode, AFL can fuzz a target without access to its source code. The necessary coverage information is collected using QEMU. According to the AFL documentation\footnote{\url{https://github.com/mirrorer/afl/blob/master/qemu\_mode/README.qemu}}, the performance may decrease substantially due to the overhead introduced by the binary instrumentation.

\noindent
\ding{186}~\textbf{AFL-CF.} As described in Section~\ref{sec:background}, AFL ships with various compilers that add the coverage feedback instrumentation when compiling a target program from source code. Using the alternative compiler \texttt{afl-clang-fast}, the instrumentation is added on the compiler level, instead of the assembly level, using a LLVM pass which improves the performance.

\noindent
\ding{187}~\textbf{AFL-LF.} Based on \texttt{afl-clang-fast}, one can try to improve the code coverage by using the LAF LLVM passes\footnote{\url{https://gitlab.com/laf-intel/laf-llvm-pass/tree/master}}. For instance, these passes split multi-byte comparisons into smaller ones which AFL can solve consecutively.

\noindent
\ding{183}~\textbf{AFLFast.} AFLFast~\cite{BOE2016} is a fork of AFL that investigates fuzzing \emph{low-frequency paths}. These are paths that are reached by only a few inputs following the intuition that these inputs solve a path constraint that may lead to a bug. The implementation is part of the power schedule with an optimized search strategy. Note that AFL incorporated improvements from AFLFast starting with version 2.31b.

\noindent
\ding{184}~\textbf{Fairfuzz.}~Fairfuzz~\cite{LEM2018} is also based on AFL. Similar to AFLFast, it aims at triggering branches that are rarely reached by other testing inputs. However, it does not utilize a Markov chain model but rather relies on a dynamic cutoff value (i.e., a threshold for the number of hits) to decide which branches are considered \emph{rare}. Further, Fairfuzz uses a heuristic that checks if certain bytes can be modified while still executing the same respective rare branch. Fairfuzz implements these changes as part of the search strategy and the mutation stage of AFL.

\noindent
\ding{185}~\textbf{AFL++.}~The AFL++ fuzzer~\cite{FIO2020} is a novel variation of AFL that improves usability and enables easy customization. It implements various improvements from academia as well as the fuzzing community (e.g., the AFLFast power schedules and the LAF LLVM passes). The goal is to introduce a new baseline fuzzer that is used for future fuzzing evaluations.

\subsection{Seed Set}
\label{subsec:experiment:seeds}

First, we evaluate the influence of the seed set by comparing an empty with a non-empty seed set (see Section~\ref{subsec:methodology:seeds}). The results are depicted in Table~\ref{tab:seed_compare} which lists the number of times that a fuzzer performed statistically better with the empty and non-empty seed set. We find that with the majority of targets the non-empty seed set either outperforms the empty seed or performs equally well on both statistical tests. We find that AFL is able to detect five bugs using the empty seed set that AFL is unable to detect when using the non-empty seed set. We believe that the main reason for this is that AFL spends less time in the deterministic stage when using an empty seed as the file is only a single byte. Note that even though the performance with a proper seed set is significantly better, testing an empty seed is still useful in cases where it is favorable to minimize the number of variables which may influence the fuzzing process~\cite{KLE2018} as well as scenarios where one cannot compile a comprehensive sets of inputs for the tested programs.

\begin{table}[htb]
\center
\small
\caption{Comparison of the non-empty and empty seed sets using interval-scale and dichotomous tests. Listed are the number of times the performance of the non-empty seed set was statistically better than the empty seed set and vice versa.}
\resizebox{0.48\textwidth}{!}{
\begin{tabular}{@{\extracolsep{3pt}}lcccc}
\hline
& \multicolumn{2}{c}{interval-scaled} & \multicolumn{2}{c}{dichotomous} \\
\cline{2-3} \cline{4-5}
Fuzzer & non-empty & empty & non-empty & empty \\
\hline
afl & 12 & 6 & 6 & 2 \\
afl (-Q) & 8 & 4 & 7 & 1 \\
afl (-d) & 18 & 2 & 8 & 1 \\
fairfuzz & 13 & 4 & 7 & 1 \\
afl-li & 13 & 6 & 5 & 3 \\
afl-cf & 12 & 6 & 5 & 2 \\
aflfast & 12 & 5 & 6 & 0 \\
afl++ & 12 & 5 & 5 & 2 \\
\hline
\end{tabular}}
\label{tab:seed_compare}
\end{table}

\subsection{Run-Time}
\label{subsec:experiment:runtime}

\begin{figure}[htb]
\centering
\begin{subfigure}{.48\textwidth}
  \centering
  \includegraphics[width=\textwidth]{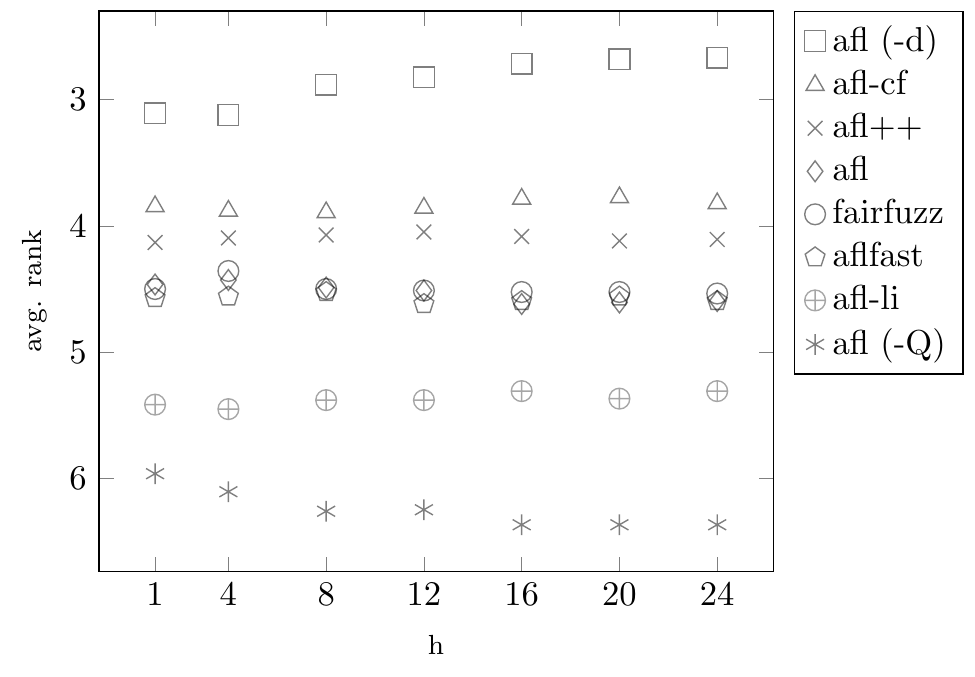}
  \caption{Run-times varying between 1h and 24h.}
  \label{fig:timeoverview}
\end{subfigure}%
\begin{subfigure}{.48\textwidth}
  \centering
  \includegraphics[width=\textwidth]{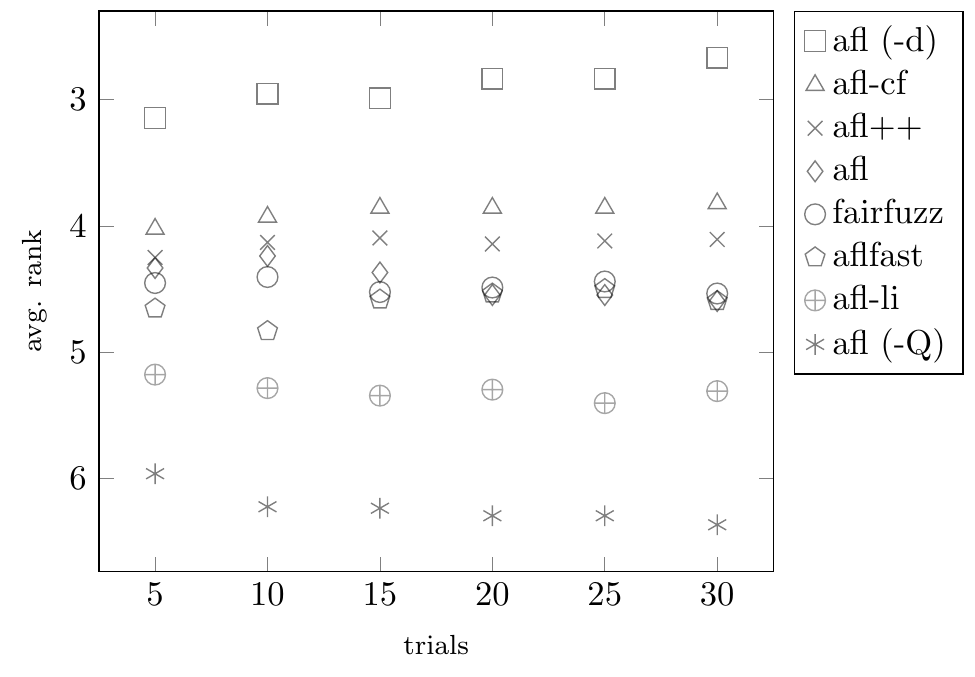}
  \caption{Trials varying between 5 and 30.}
  \label{fig:repoverview2}
\end{subfigure}
\caption{Average ranking when using different run-times and number of trials.}
\label{fig:results}
\end{figure}

To show the impact of differences in run-time, we calculate the ranking for maximum run-times between 1h and 24h. For each ranking, we only consider crashes that have been found in the respective time frame. We present the results in Figure~\ref{fig:timeoverview}. We observe that the run-time has a significant influence on the results. Interestingly, we find that even though all fuzzers are based on the same code base there is no uniform trend when increasing the run-time. For example, AFL without its deterministic stage consistently improves, in total by 0.45 in the average ranking from 1h to 24h. In the same time the performance of Fairfuzz, AFLFast, and AFL may increase or decrease slightly which also changes the relative ranking of these fuzzers depending on the maximum run-time. \emph{We observe that on our test set, the run-time should be at least 8h as lower run-times may lead to false conclusions of the fuzzer performance.}

\subsection{Number of Trials}
\label{subsec:experiment:trials}

To calculate a p-value, one has to repeat every experiment multiple times. The number of trials also influences the minimum p-value that can be achieved. We compare the average ranking of each fuzzer and configuration considering between the first 5 and all 30 trials. In Figure~\ref{fig:repoverview2}, we can see that the performance may vary significantly depending on the number of trials used. For example, using 10 trials AFL++ has a slightly better performance than AFL and Fairfuzz, both of which clearly outperform AFLFast. Analyzing the results after 30 trials we find that AFL++ now outperforms AFL and Fairfuzz which both perform as well as AFLFast. \emph{We conclude that the number of trials has significant impact on the results and if under serious resource constraints one should prioritize a higher number of trials over longer run-times.}

\subsection{Number of Bugs/Targets}
\label{subsec:experiment:bugs}

\begin{figure*}[htb]
\center
\makebox[\textwidth][c]{\includegraphics[width=1.3\textwidth]{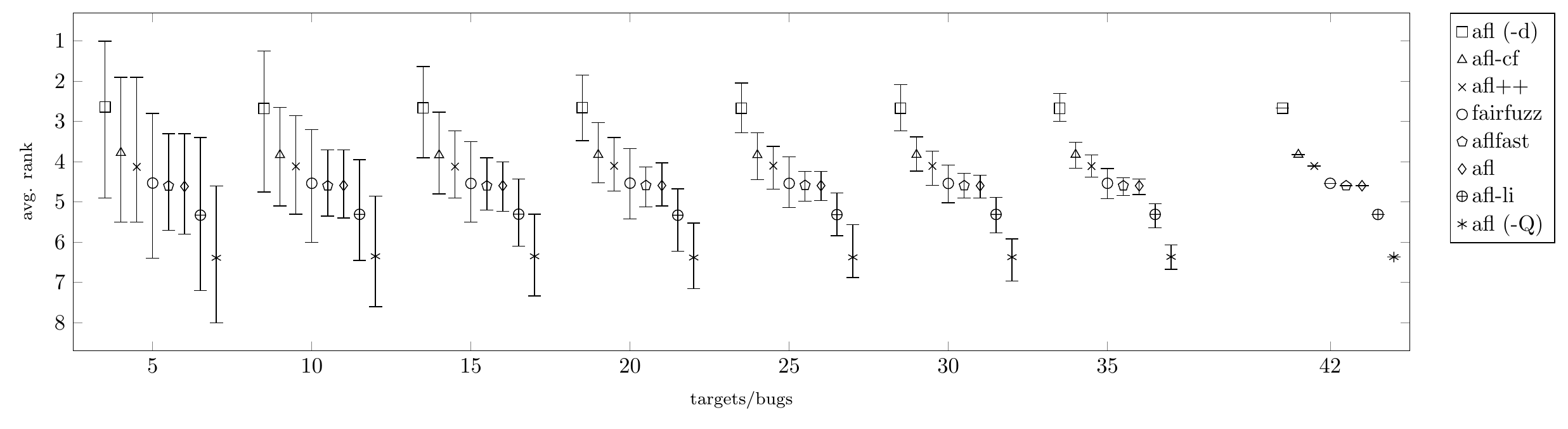}}
\caption{Average ranking when using varying numbers of targets/bugs. Whiskers correlate to the minimum and maximum rank.}
\label{fig:targetsoverview}
\end{figure*}

Another parameter that one can adjust is the number of targets a fuzzer is evaluated on. As we use targets with a single bug, the number of targets is equal to the number of bugs in our test set. We evaluate all fuzzers on test sets between five and 35 different targets. For each test set size, we randomly sample 1000 different target combinations and calculate the ranking including maximum and minimum. Note that given larger test sets, the spread will naturally decrease as we sample from a maximum of 42 different targets. In Figure~\ref{fig:targetsoverview}, we can see that the performance may vary substantially depending on the used test set. We further analyze the results and find randomly sampled test sets with 15 targets where AFL-CF outperforms AFL without the deterministic stage or test sets where the performance of Fairfuzz is second to last. Even when we use 35 targets, we find randomly sampled test sets that result in a substantially different rankings compared to the 42 target test set. For example, we observe test sets where AFL++ outperforms AFL-CF or test sets where Fairfuzz performs better than AFL++. \emph{Our results show that target and bug selection should not be taken lightly as it can introduce significant biases when testing fuzzers.}

\subsection{Statistical Significance}
\label{subsec:experiments:statistical_sig}

\begin{figure}[htb]
\centering
\begin{subfigure}{.48\textwidth}
  \centering
  \includegraphics[width=\textwidth]{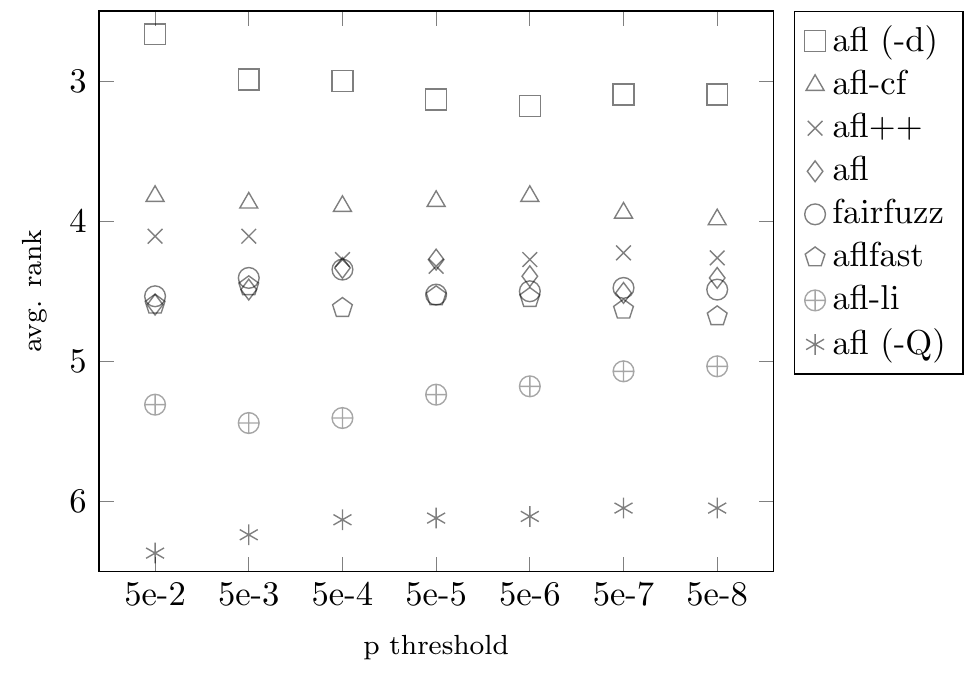}
  \caption{Interval-scaled ranking}
  \label{fig:p}
\end{subfigure}%
\begin{subfigure}{.48\textwidth}
  \centering
  \includegraphics[width=\textwidth]{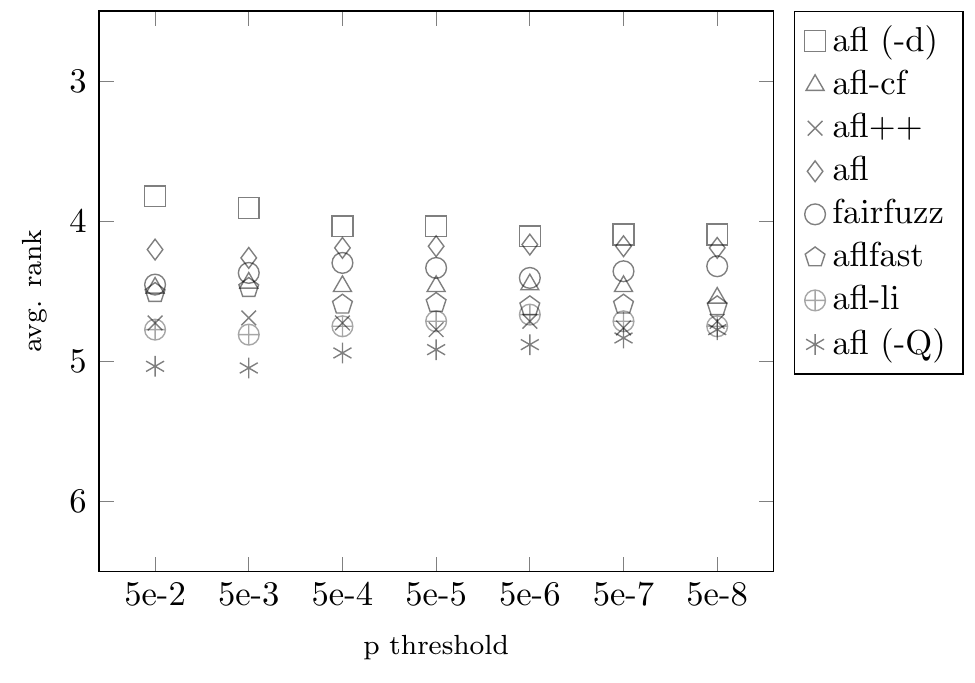}
  \caption{Dichotomous ranking.}
  \label{fig:p_dicho}
\end{subfigure}
\caption{Comparison of the average rank for various p thresholds.}
\label{fig:p_thresholds}
\end{figure}

Recall that \framework\ supports two different statistical tests: dichotomous and interval-scaled (see Section~\ref{subsec:backgroud:statisticalevaluation}). In Figure~\ref{fig:p_thresholds} we compare the average ranking resulting from both methods. Looking at the results for 5e-2, we can see that the ranking based on dichotomous tests shows less performance differences. Interestingly, we find that AFL++ is performing relatively worse on the dichotomous ranking, as it ranks slightly below AFL-CF, Fairfuzz, and AFLFast. Futhermore, AFL is the second best performing fuzzer, while AFL without the deterministic stage is still the best performing fuzzers. \emph{The difference of both ranking shows that it is useful to provide both, inverval-scaled and dichotomous results as they test different aspects of fuzzers.}

Klees et al.~\cite{KLE2018} as well as Arcuri and Briand~\cite{ARC2014} recommend to use a p~threshold of 0.05. As other scientific communities opted to use lower thresholds~\cite{BEN2017} we analyze the influence of lower p~thresholds. Generally, lowering the p~threshold decreases the chances for false positive results while increasing the chance for false negatives. We calculate the ranking for each threshold between 0.05 and 5e-8 and provide the results in Figure~\ref{fig:p_thresholds}. We can see that the interval-scaled ranking is less affected when using a lower threshold compared to the dichotomous ranking. Taking a closer look at Figure~\ref{fig:p}, we observe that the relative ranking of AFLFast, Fairfuzz, and AFL is affected the most due to the similar performance of all three fuzzers. Interestingly, we find that a lower p threshold greatly influences the relative performance of AFL++. While clearly in third place with a p~threshold of 0.05, AFL++ performs on a similar level as AFL when using a threshold of 5e-4 to 5e-6. \emph{Thus, we can see that the p threshold can have a significant effect when comparing the performance of fuzzers.}

\subsection{Effect Size}
\label{subsec:experiments:effect_size}

\begin{figure}[htb]
\center
\includegraphics[width=0.48\columnwidth]{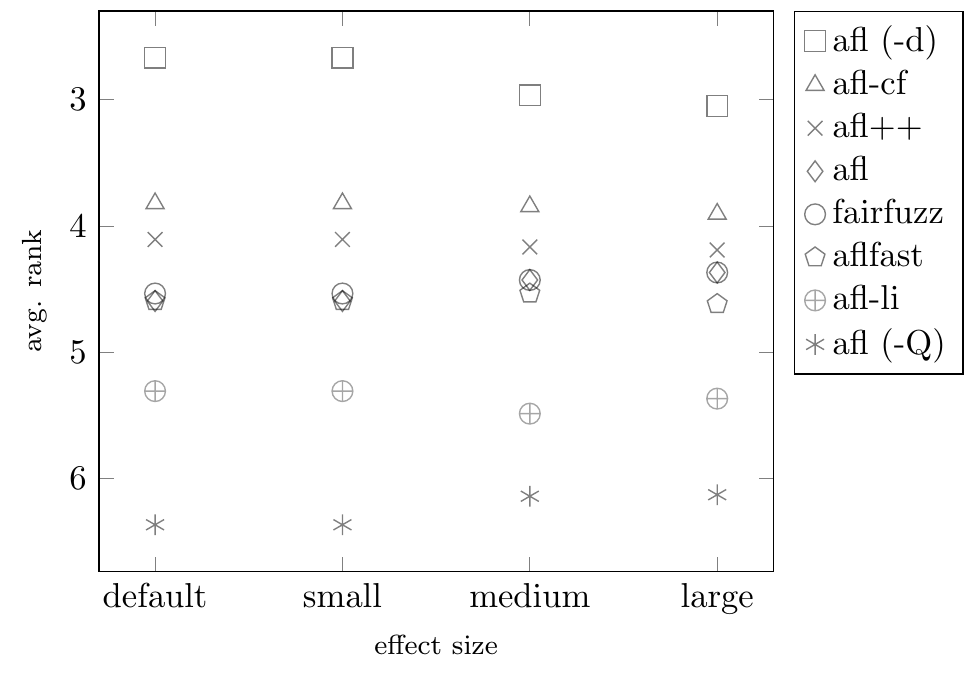}
\caption{Difference between different effect size tresholds when using the $\hat{A}_{12}$ statistic.}
\label{fig:a12_diff}
\end{figure}

Besides the threshold for statistical significance, one can also set a threshold for the effect size. Vargha and Delaney~\cite{VAR00} provide three different thresholds for the $\hat{A}_{12}$ effect size. Namely 0.56 is considered small, 0.64 is considered medium and 0.71 is called a large effect. We provide an overview of the average ranking when considering all or only small/medium/large effect sizes in Figure~\ref{fig:a12_diff}. We can observe that the influence of the effect size is rather subtle compared to the other parameters. The mostly effected fuzzers are, AFL++, Fairfuzz, AFLFast, and AFL. Most notably, the performance of AFLFast is relatively worse when only considering large differences. \emph{We conclude that the effect size threshold is less meaningful when comparing fuzzers in our evaluation setup.}

\subsection{Further Insights}
\label{subsec:experiments:furtherinsights}

Next, we compare the \framework-ranking with a ranking that utilizes the \emph{average} as commonly found in fuzzer evaluations (see Section~\ref{sec:problem_description}). The results are shown in Table~\ref{tab:avg}. Notably, when we only consider the average, the overall ranking changes drastically with the exception of the best and worst performing fuzzers. \emph{This shows the influence of statistically insignificant results on the overall performance results which further confirms the choice of using righteous statistical methods as employed in \framework.}

\begin{table}[htb]
\center
\small
\caption{Comparison of the \framework-ranking and avg. number of bugs found over all targets.}
\resizebox{0.48\textwidth}{!}{
\begin{tabular}{@{\extracolsep{3pt}}lrlr}
\hline
\multicolumn{2}{c}{SENF Ranking} &\multicolumn{2}{c}{Ranking based on Avg.} \\
\cline{1-2} \cline{3-4}
Fuzzer & Avg. ranking & Fuzzer & Avg. \#bugs found\\
\hline
afl (-d) & 2.67 & afl (-d) & 16.86 \\
afl-cf & 3.82 & fairfuzz	& 14.17\\ 
afl++ & 4.11 & afl-cf & 13.55 \\
fairfuzz & 4.54 & aflfast & 13.26\\ 
aflfast & 4.60 & afl & 12.60 \\ 
afl & 4.60 & afl-li & 12.60 \\
afl-li & 5.31 &  afl++ & 12.48\\
afl (-Q) & 6.37 & afl (-Q) & 9.21\\
\hline
\end{tabular}}
\label{tab:avg}
\end{table}

\begin{figure}[htb]
\center
\includegraphics[width=0.48\columnwidth]{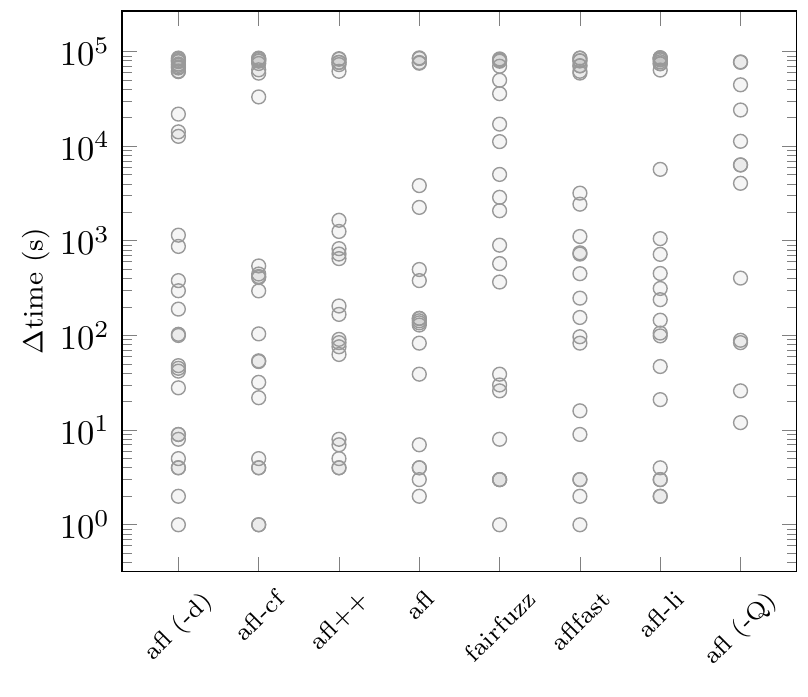}
\caption{Difference between min. and max. exec. time for each fuzzer and target over all trials.}
\label{fig:execdiff}
\end{figure}

To test the \emph{consistency} of each fuzzer, we take a closer look at the time it takes a fuzzer to detect a bug in Figure~\ref{fig:execdiff}. To improve the readability of the figure we plot the difference between the shortest and longest time a fuzzer needs to find a bug over all trials for each target. If a fuzzer is not able to find a bug, we set the execution time to 24h. When a fuzzer was not able to find a bug in a target over all trials, we omitted the result to increase readability. For all fuzzers and configurations, randomness plays a significant role when searching for bugs with differences between minimum and maximum time close to our run-time of 24h. \emph{No fuzzer in our evaluation is able to consistently find bugs over all trials.}

%% file: 07-discussion_extended.tex
\section{Discussion}
\label{sec:discussion}

\noindent
\textbf{Test Set Selection.}
Our framework \framework\ in combination with a ground truth test set significantly increases the probability that the reported results are reproducible. Even though our test set of 42 different programs and 21 different bug types ensures a certain level of diversity in our evaluation, the resulting ranking could potentially differ if a larger, representative test set of real-world programs with a ground truth is used because programs from the CGC test set do not provide the same level complexity. Note that other test sets can easily be evaluated with \framework\ as we only require a database containing the experiment results as input.

\medskip
\noindent
\textbf{Resource Limitations.}
Due to unavoidable limitations of resources, we cannot analyze the full range of parameters used in existing fuzzing evaluations, e.g., run-times of 60h (see Section~\ref{sec:problem_description}). Therefore, we limit our experiments to values recommended in fuzzing literature~\cite{KLE2018}. For the same reason, we do not conduct experiments with multiple concurrent fuzzer instances testing the same target. The experiments of Chen et al.~\cite{CHE2019} as well as B\"{o}hme and Falk~\cite{BOE2020} suggest that the performance of fuzzers varies significantly when fuzzing with multiple instances simultaneously.

\medskip
\noindent
\textbf{Fuzzer Selection.}
Due to the aforementioned resource constraints, we have to limit the selection of fuzzers as the experiments in Section~\ref{sec:experiments} already required over 280k CPU hours. We opted to focus on AFL, AFL-based fuzzers, and various optimizations as this allows us to easily attribute performance differences. Furthermore, AFL is the most popular baseline fuzzer, e.g., it is recommended by Klees et al.~\cite{KLE2018} and used in all evaluations we studied in Section~\ref{sec:problem_description}. Additionally, AFL is commonly used as a code base to implement new fuzzers~\cite{BOE2016,SCH2017,GAN2018,LEM2018,LYU2019,PHA2019}. For these reasons, we argue that focusing on AFL-style fuzzers is more significant for common fuzzer evaluations compared to other fuzzers. However, since our implementation is open-source one can easily use \framework\ to evaluate other fuzzers. We provide detailed guidelines in our public repository.

\medskip
\noindent
\textbf{Scoring Algorithm.}
The scoring algorithm we use in our evaluation adopts the commonly used intuition that the fuzzer which outperforms the other fuzzers (i.e., finds more bugs) on the most targets has the best overall performance. However, other evaluation metrics may be useful for other use cases, e.g., when testing a target with a section of different fuzzers one may not only be interested in the fuzzer that finds the most bugs but also fuzzers that find a unique set bugs which all other fuzzers are unable to detect. However, calculating the unique set of bugs for each fuzzer does not require complex statistical evaluations as provided by \framework.

Furthermore, our evaluation does not take into account by how much a fuzzer~A improves over a different fuzzer~B. \framework\ addresses this problem by supporting a variable effect size thresholds. Thus, interested parties can set a custom minimum effect size which \framework\ takes into account when calculating the score of a fuzzer. We provide more detailed information on the effect size and its influence on the fuzzer evaluation in Section~\ref{subsec:experiments:effect_size}.

\medskip
\noindent
\textbf{Threshold of the p-value.}
In our evaluation, we opted to use the widely established p threshold of 0.05 which is commonly used in software engineering evaluations~\cite{KLE2018}. However, this threshold is generally considered a trade-off between the probability of false positive and false negative results. Other scientific communities opted to use lower thresholds or other methods of statistical evaluation~\cite{BEN2017}. \framework\ addresses this and lets the user set an arbitrary threshold to calculate the average ranking of each fuzzer. A more detailed discussion on the influence of the p-value threshold is given in Section~\ref{subsec:experiments:statistical_sig}.

%% file: 08-conclusion.tex
\section{Conclusion}
\label{sec:conclusion}

Our analysis of recent fuzzing studies shows that fuzzers are largely evaluated with various different evaluation parameters which are not in line with the recommendations found in academic literature. To address these issues, we presented \framework, which implements dichotomous and interval-scale statistical methods to calculate the p-value and effect sizes to compute a ranking to asses the overall performance of all tested fuzzers.

Based on extensive empirical data, we quantified the influence of different evaluation parameters on fuzzing evaluations for the first time. We demonstrate that even when we utilize the recommended statistical tests, using insufficient evaluation parameters --- such as a low number of trials or a small test set --- may still lead to misleading results that in turn may lead to false conclusions about the performance of a fuzzer. Thus, the choice of parameters for fuzzing evaluations should not be taken lightly and existing recommendations should be followed to lower the chance of non-reproducible results. We described and open-sourced a practical evaluation setup that can be used to test the performance of fuzzers.

%% file: 09-acknowledgements.tex
\section*{Acknowledgements}

Funded by the Deutsche Forschungsgemeinschaft (DFG, German Research Foundation) under Germany's Excellence Strategy -- EXC 2092 CASA -- 390781972.

\noindent
This work has been partially funded by the Deutsche Forschungsgemeinschaft (DFG, German Research Foundation) -- SFB 1119 -- 236615297.

%% file: 10-appendix.tex
\appendix
\section{Test set}
\label{apx:subsec:testset}
In Table~\ref{tab:apx:targets} we provide a list with all targets in our test set including the CWE classification of the included bug that can be found by a fuzzer. Note that seven targets cannot be included when using an empty seed due to requirements of AFL.

\begin{table}[!hb]
\centering
\small
\caption{List of all target names from the CGC test set ported by Trail of Bits included in our test set. For each bug, we provide the corresponding CWE classification found in the respective challenge description.}
\begin{threeparttable}
\begin{tabular}{rp{0.4\columnwidth}p{0.4\columnwidth}}
\hline
Alias & Name & CWE classes\\ 
\hline 
01 & AIS-Lite & CWE-20, CWE-120, CWE-122, CWE-129, CWE-788\\ 
02 & basic\_emulator & CWE-170\\ 
03 & BitBlaster & CWE-476, CWE-824 \\ 
04 & Cereal\_Mixup\_\_A\_Cereal\_\allowbreak Vending\_\allowbreak Machine\_\allowbreak Controller & CWE-502, CWE-822\\ 
05 & CNMP & CWE-134\\ 
06 & Differ\tnote{1} & CWE-121\\
07 & Diophantine\_\allowbreak Password\_\allowbreak Wallet & CWE-476, CWE-824 \\ 
08 & Divelogger2\tnote{1} & CWE-119\\ 
09& expression\_database\tnote{1} & CWE-119\\
10 & FileSys & CWE-416\\ 
11 & FSK\_Messaging\_Service & CWE-120, CWE-122 \\
12 & humaninterface & CWE-122\\ 
13 & Image\_Compressor & CWE-787\\ 
14 & LazyCalc & CWE-121\\ 
15 & Loud\_Square\_\allowbreak Instant\_\allowbreak Messaging\_\allowbreak Protocol\_\allowbreak LSIMP & CWE-843\\ 
16 & Matrix\_Math\_Calculator & CWE-121\\ 
17 & middleout & CWE-121, CWE-788, CWE-787\\
18 & Movie\_Rental\tnote{1} & CWE-134\\ 
19 & Multi\_User\_Calendar & CWE-121\\ 
20 & Multipass & CWE-822\\ 
21 & One\_Amp & CWE-121\\
22 & online\_job\_application2\tnote{1} & CWE-120, CWE-122\\ 
23 & Palindrome & CWE-121\\ 
24 & Palindrome2 & CWE-121\\ 
25 & Particle\_Simulator & CWE-787\\ 
26 & Personal\_Fitness\_Manager & CWE-121, CWE-131\\ 
27 & Printer & CWE-122\\ 
28 & REMATCH\_3--Address\_Resolution\_Service--SQL\_Slammer & CWE-121\\ 
29 & Resort\_Modeller & CWE-468, CWE-822\\ 
30 & root64\_and\_parcour & CWE-121\\ 
31 & SAuth & CWE-121, CWE-190\\ 
32 & SFTSCBSISS & CWE-120\\ 
33 & ShoutCTF & CWE-121\\ 
34 & Simple\_Stack\_Machine & CWE-122\\ 
35 & Space\_Attackers &	CWE-121\\ 
36 & Square\_Rabbit & CWE-190\\ 
37 & stream\_vm\tnote{1} & CWE-127\\ 
38 & stream\_vm2 & CWE-129, CWE-665, CWE-787\\ 
39 & Thermal\_Controller\_v2 & CWE-121\\ 
40 & User\_Manager\tnote{1} & CWE-416\\
41 & XStore & CWE-121\\ 
42 & yolodex & CWE-787\\
\hline
\end{tabular}
\begin{tablenotes}\footnotesize
\item[1] Target cannot process an empty seed.
\end{tablenotes}
\end{threeparttable}
\label{tab:apx:targets}
\end{table}

%% file: main.bbl
\begin{thebibliography}{10}
\providecommand{\url}[1]{\texttt{#1}}
\providecommand{\urlprefix}{URL }
\providecommand{\doi}[1]{https://doi.org/#1}

\bibitem{AIZ2016}
Aizatsky, M., Serebryany, K., Chang, O., Arya, A., Whittaker, M.: {Announcing
  OSS-Fuzz: Continuous Fuzzing for Open Source Software} (2016)

\bibitem{ARC2014}
Arcuri, A., Briand, L.: {A Hitchhiker's guide to statistical tests for
  assessing randomized algorithms in software engineering}. Software Testing,
  Verification and Reliability  (2014)

\bibitem{ASH2019}
Aschermann, C., Schumilo, S., Blazytko, T., Gawlik, R., Holz, T.: {REDQUEEN:
  Fuzzing with Input-to-State Correspondence}. In: Symposium on Network and
  Distributed System Security (NDSS) (2019)

\bibitem{BEN2017}
Benjamin, D.J., Berger, J.O., Johannesson, M., Nosek, B.A., Wagenmakers, E.J.,
  et~al.: {Redefine Statistical Significance}. Human Nature Behavior  (2017)

\bibitem{BLA2019}
Blazytko, T., Aschermann, C., Schl\"{o}gel, M., Abbasi, A., Schumilo, S.,
  W\"{o}rner, S., Holz, T.: {GRIMOIRE: Synthesizing Structure While Fuzzing}.
  In: USENIX Security Symposium (2019)

\bibitem{BOE2020}
B\"{o}hme, M., Falk, B.: Fuzzing: On the exponential cost of vulnerability
  discovery. In: Symposium on the Foundations of Software Engineering (FSE)
  (2020)

\bibitem{BOE2017}
B\"{o}hme, M., Pham, V.T., Nguyen, M.D., Roychoudhury, A.: Directed greybox
  fuzzing. In: ACM Conference on Computer and Communications Security (CCS)
  (2017)

\bibitem{BOE2016}
B\"{o}hme, M., Pham, V.T., Roychoudhury, A.: {Coverage-Based Greybox Fuzzing as
  Markov Chain}. In: ACM Conference on Computer and Communications Security
  (CCS) (2016)

\bibitem{CAD2008}
Cadar, C., Dunbar, D., Engler, D.: {KLEE: Unassisted and Automatic Generation
  of High-Coverage Tests for Complex Systems Programs}. In: USENIX Conference
  on Operating Systems Design and Implementation (2008)

\bibitem{HON2018}
Chen, H., Xue, Y., Li, Y., Chen, B., Xie, X., Wu, X., Liu, Y.: {Hawkeye:
  Towards a Desired Directed Grey-Box Fuzzer}. In: ACM Conference on Computer
  and Communications Security (CCS) (2018)

\bibitem{CHE2018}
Chen, P., Chen, H.: {Angora: Efficient Fuzzing by Principled Search}. In: IEEE
  Symposium on Security and Privacy (S\&P) (2018)

\bibitem{CHE2019}
Chen, Y., Jiang, Y., Ma, F., Liang, J., Wang, M., Zhou, C., Jiao, X., Su, Z.:
  {EnFuzz: Ensemble Fuzzing with Seed Synchronization among Diverse Fuzzers}.
  In: USENIX Security Symposium (2019)

\bibitem{CHO2019}
Cho, M., Kim, S., Kwon, T.: {Intriguer: Field-Level Constraint Solving for
  Hybrid Fuzzing}. In: ACM Conference on Computer and Communications Security
  (CCS) (2019)

\bibitem{DOL2016}
{Dolan-Gavitt}, B., Hulin, P., Kirda, E., Leek, T., Mambretti, A., Robertson,
  W., Ulrich, F., Whelan, R.: {LAVA: Large-Scale Automated Vulnerability
  Addition}. In: IEEE Symposium on Security and Privacy (S\&P) (2016)

\bibitem{FIO2020}
Fioraldi, A., Maier, D., Ei{\ss}feldt, H., Heuse, M.: Afl++ : Combining
  incremental steps of fuzzing research. In: {USENIX} Workshop on Offensive
  Technologies ({WOOT}) (2020)

\bibitem{FIS22}
Fisher, R.: {On the Interpretation of {$\chi^2$} from Contingency Tables, and
  the Calculation of P}. Journal of the Royal Statistical Society  \textbf{85}
  (1922)

\bibitem{FIS1925}
Fisher, R.: {{Statistical Methods for Research Workers}}. Oliver and Boyd
  (1925)

\bibitem{GAN2020}
Gan, S., Zhang, C., Chen, P., Zhao, B., Qin, X., Wu, D., Chen, Z.: {GREYONE}:
  Data flow sensitive fuzzing. In: USENIX Security Symposium (2020)

\bibitem{GAN2018}
Gan, S., Zhang, C., Qin, X., Tu, X., Li, K., Pei, Z., Chen, Zhuoning, C.:
  {CollAFL: Path Sensitive Fuzzing}. In: IEEE Symposium on Security and Privacy
  (S\&P) (2018)

\bibitem{GOO2016}
Google: {fuzzer-test-suite}. \url{https://github.com/google/fuzzer-test-suite/}
  (2016)

\bibitem{GUI2016}
Guido, D.: {Your tool works better than mine? Prove it.}
  \url{https://blog.trailofbits.com/2016/08/01/your-tool-works-better-than-mine-prove-it/}
  (2016)

\bibitem{HAZ2020}
Hazimeh, A., Herrera, A., Payer, M.: Magma: A ground-truth fuzzing benchmark.
  Proceedings of the ACM on Measurement and Analysis of Computing Systems
  \textbf{4} (2020)

\bibitem{HOV2006}
Hocevar, S.: {zzuf}. \url{https://github.com/samhocevar/zzuf/} (2006)

\bibitem{HUA2020}
Huang, H., Yao, P., Wu, R., Shi, Q., Zhang, C.: Pangolin: Incremental hybrid
  fuzzing with polyhedral path abstraction. In: IEEE Symposium on Security and
  Privacy (S\&P) (2020)

\bibitem{INO2014}
Inozemtseva, L., Holmes, R.: Coverage is not strongly correlated with test
  suite effectiveness. In: International Conference on Software Engineering
  (ICSE) (2014)

\bibitem{KLE2018}
Klees, G., Ruef, A., Cooper, B., Wei, S., Hicks, M.: {Evaluating Fuzz Testing}.
  In: ACM Conference on Computer and Communications Security (CCS) (2018)

\bibitem{LEM2018}
Lemieux, C., Sen, K.: {FairFuzz: A Targeted Mutation Strategy for Increasing
  Greybox Fuzz Testing Coverage}  (2018)

\bibitem{LI2021}
Li, Y., Ji, S., Chen, Y., Liang, S., Lee, W.H., Chen, Y., Lyu, C., Wu, C.,
  Beyah, R., Cheng, P., Lu, K., Wang, T.: {UNIFUZZ}: A holistic and pragmatic
  metrics-driven platform for evaluating fuzzers. In: USENIX Security Symposium
  (2021)

\bibitem{LYU2019}
Lyu, C., Ji, S., Zhang, C., Li, Y., Lee, W.H., Song, Y., Beyah, R.: {MOPT:
  Optimized Mutation Scheduling for Fuzzers}. In: USENIX Security Symposium
  (2019)

\bibitem{MAN47}
Mann, H., Whitney, D.: {On a Test of Whether one of Two Random Variables is
  Stochastically Larger than the Other}. Annals of Mathematical Statistics
  \textbf{18} (1947)

\bibitem{MET2020}
Metzman, J., Arya, A., Szekeres, L.: {FuzzBench: Fuzzer Benchmarking as a
  Service}.
  \url{https://opensource.googleblog.com/2020/03/fuzzbench-fuzzer-benchmarking-as-service.html}
  (2020)

\bibitem{PAA2021}
Paa{\ss}en, D., Surminski, S., Rodler, M., Davi, L.: {Public github respository
  of SENF}. \url{https://github.com/uni-due-syssec/SENF}

\bibitem{PHA2019}
Pham, V.T., B{\"o}hme, M., Santosa, A.E., C\u{a}ciulescu, A.R., Roychoudhury,
  A.: Smart greybox fuzzing. IEEE Transactions on Software Engineering  (2019)

\bibitem{PHA2017}
Pham, V.T., Khurana, S., Roy, S., Roychoudhury, A.: {Bucketing Failing Tests
  via Symbolic Analysis}. In: International Conference on Fundamental
  Approaches to Software Engineering (2017)

\bibitem{RCT2019}
{R Core Team}: R: A Language and Environment for Statistical Computing. R
  Foundation for Statistical Computing, Vienna, Austria (2019),
  \url{https://www.R-project.org/}

\bibitem{SCH2017}
Schumilo, S., Aschermann, C., Gawlik, R., Schinzel, S., Holz, T.: {kAFL:
  Hardware-Assisted Feedback Fuzzing for OS Kernels}. In: USENIX Security
  Symposium (2017)

\bibitem{VTO2018}
van Tonder, R., Kotheimer, J., Le~Goues, C.: {Semantic Crash Bucketing}. In:
  ACM/IEEE International Conference on Automated Software Engineering (2018)

\bibitem{VAR00}
Vargha, A., Delaney, H.D.: {A Critique and Improvement of the CL Common
  Language Effect Size Statistics of McGraw and Wong}. Journal of Educational
  and Behavioral Statistics  \textbf{25} (2000)

\bibitem{WAN2020}
Wang, Y., Jia, X., Liu, Y., Zeng, K., Bao, T., Wu, D., Su, P.: Not all coverage
  measurements are equal: Fuzzing for input prioritization. In: Symposium on
  Network and Distributed System Security (NDSS) (2020)

\bibitem{YUE2020}
Yue, T., Wang, P., Tang, Y., Wang, E., Yu, B., Lu, K., Zhou, X.: Ecofuzz:
  Adaptive energy-saving greybox fuzzing as a variant of the adversarial
  multi-armed bandit. In: USENIX Security Symposium (2020)

\bibitem{ZAL2019}
Zalewski, M.: {Technical "whitepaper" for afl-fuzz}.
  \url{https://lcamtuf.coredump.cx/afl/technical\_details.txt}

\bibitem{ZHA2019}
Zhao, L., Duan, Y., Yin, H., Xuan, J.: {Send Hardest Problems My Way:
  Probabilistic Path Prioritization for Hybrid Fuzzing}. In: Symposium on
  Network and Distributed System Security (NDSS) (2019)

\end{thebibliography}
